\documentclass[aps,preprint,showpacs,preprintnumbers,nofootinbib]{revtex4}
\usepackage{graphicx,,color,amsmath}
\begin{document}
\title{Constraints on String Resonance Amplitudes}
\author{Kingman Cheung and Yueh-Feng Liu}
\email[E-mail: ]{cheung@phys.nthu.edu.tw, g923310@oz.nthu.edu.tw}
 \affiliation{Department of Physics and
NCTS, National Tsing Hua University, Hsinchu, Taiwan R.O.C}
\date{\today}

\begin{abstract}
We perform a global analysis of the tree-level open-string
amplitudes in the limit $s \ll M_S^2$ . Based on the present data
from the Tevatron, HERA, and LEP~2, we set a lower limit on the
string scale $M_S \geq 0.69 - 1.96$ TeV at 95\% confidence level
for the Chan-Paton factors $|T|=0 - 4$. We also estimate the
expected sensitivities at the LHC, which can be as high as 19
TeV for $|T|=4$.
\end{abstract}
\maketitle

\section{Introduction}
String theory~\cite{witten} provides a consistent framework for
the unification of gravity with the standard model (SM) gauge
interactions. The string scale $M_S$ is naturally close to the
quantum gravity scale $M_{pl} \approx 10^{19}$ GeV. However,
Arkani-Hamed, Dimopoulos, and Dvali~\cite{ADD} have proposed an
alternative to this scenario. They assumed the space-time is more
than 4 dimensions with the SM particles living on a D3-brane.
While the electromagnetic, strong, and weak forces are confined to
this brane, gravity can propagate in the extra dimensions. The
effective fundamental Planck scale is of the order of TeV with
large extra dimensions as large as $mm$. TeV scale string theories
are also possible with advance of D-branes~\cite{lykken,shiu}.
Thus, make it possible for experiments to test and probe the
theory at high energy colliders~\cite{K.C.}.

In the simplest open-string model~\cite{cullen,cornet,friess}, it
is assumed that all SM particles are identified as open strings
confined to a D3-brane universe while a graviton is a closed
string propagating freely in the bulk. One novel feature of the
open-string model is the appearance of string resonances in the
scattering of particles. Earlier works on phenomenological
studies~\cite{cullen,cornet,friess,burikham,t.han} based on open-string
amplitudes have been performed, which reduce to SM tree-level
amplitudes at low energies . When $s > M_S^2$, clear resonances can
be observed in experiments, in particular the mass spectrum of the
resonances is given by $\sqrt{n}M_S(n=1,2,3...)$, which is very
different from the Kaluza-Klein states in the ADD \cite{ADD} or
Randall-Sundrum~\cite{rs} model. On the other hand, when $s <
M_S^2$ there would not appear resonances but may experience virtual
interference effects from the string resonances. Effectively, the
virtual effect is similarly to a contact interaction scaling as
$1/M_S^4$. Note that the usual 4-fermion contact interaction due
to exchange of new gauge bosons or leptoquarks has the leading
effect of $1/\Lambda^2$, where $\Lambda$ is the scale of new
physics~\cite{kingman}. Bounds on four-fermion contact
interactions induced by string resonances have been estimated
based on the analysis on $1/\Lambda^2$ contact
interaction~\cite{burikham}. Although the approximation is valid,
we shall show in this work that the effect is more complicated
than just the simple substitution $1/\Lambda^2 \rightarrow
1/M_S^4$.

The major improvement from previous work is as follows. We start
directly from open-string amplitudes and perform a global analysis
instead of using the result of an analysis on contact interaction
\cite{kccont}. The global
data include Drell-Yan production at the Tevatron, HERA charged
and neutral current deep-inelastic scattering, total hadronic, and
leptonic $e^+e^-\to e^+e^-, \mu^+\mu^-, \tau^+\tau^-$ cross
section at LEP~2. Some of the data (CDF, HERA, and LEP~2) have
been updated since the summer of 2004. We shall see that Drell-Yan
production, due to large invariant mass data, provides the
strongest constraint among the global data. By combining all data,
the string scale $M_S$ must be larger than $0.69 - 1.96$ TeV for
$|T|=0 - 4$ at 95\% confidence level (CL).

The organization of the paper is as follows. In the next section,
we describe the tree-level open-string amplitudes and its low
energy approximation. In Sec.~\ref{sec:data} we describe the high
energy data sets that we used in this analysis. In
Sec.~\ref{sec:results}, we present our results on the fits and
limits. In Sec.~\ref{sec:sensitivity}, we estimate the sensitivity
at the LHC. We summarize in Sec.~\ref{sec:summary}.

\section{Open-String Amplitudes}
Our construction of tree-level open-string amplitudes follows the
same approach as that outlined in
refs.~\cite{cornet,friess,burikham,t.han}. They assumed that the
tree-level open-string amplitudes represent the scattering of
massless SM particles as the zeroth string modes. The SM is
embedded in a type IIB string theory whose 10-dimensional space
has six dimensions compactified on a torus with a common period
$2\pi R$. There are ${\cal N}$ coincident D3-brane, on which open
strings may end, that lie in the 4 extended dimensions. This
approach assures the correct low energy phenomenology as given by
the SM, yet captures one of the essential features of string
theory, namely the string resonances.

The 4-point tree-level open-string amplitude is given as a sum of
ordered amplitudes multiplied by group theory Chan-Paton factors
and can be expressed generically as~\cite{cullen,cornet,friess}
\begin{eqnarray}
\label{strgen}
  {\cal
  A}_{\it
  string}=S(s,t)A_{1234}T_{1234}+S(t,u)A_{1324}T_{1324}+S(u,s)A_{1243}T_{1243},
\;
\end{eqnarray}
where ${A_{ijkl}}'s$ are kinematic parts of the amplitude. The
Mandelstam variables are denoted by $s, t$ and $u$. ${T_{ijkl}}'s$
are the Chan-Paton factors~\cite{cp} that involve traces over the
group representation matrices $\lambda$. For example,
$$
T_{1234}=tr(\lambda_1\lambda_2\lambda_3\lambda_4)+tr(\lambda_4\lambda_3
\lambda_2\lambda_1).
$$
Typically, with normalization $Tr(\lambda^a\lambda^b)=\delta^{ab}$
the Chan-Paton factors are in the range of $-4$ to 4. Since a
complete string model construction for the electroweak interaction
of the SM is unavailable, Chan-Paton factors are taken as free
parameters in our calculation. $S(s,t)$ is essentially the
Veneziano amplitude~\cite{venz}
\begin{eqnarray}
S(s,t)=\frac{\Gamma(1-\alpha's)\Gamma(1-\alpha't)}
{\Gamma(1-\alpha's-\alpha't)}\; ,
\end{eqnarray}
where the Regge slope $\alpha'=M_S^{-2}$, and $S\rightarrow 1$ as
either $s/M_S^2$ or $t/M_S^2\rightarrow 0$.

We start with Drell-Yan production of a pair of leptons. In the
massless limit of the fermions we label their chirality
$\alpha,\beta=L,R$. The tree-level open-string amplitude is
\begin{eqnarray}
\label{str}
 {\cal A}_{\it string}(q_L\overline{q_L}\rightarrow \ell_R\overline{
\ell_R})=ig^2_S\left[T_{1234}S(s,t)\frac{t}{s}+T_{1324}S(t,u)\frac{t}{u}+
T_{1243}S(u,s)(-\frac{t}{s}-\frac{t}{u})\right]\;.
\end{eqnarray}
The corresponding standard model amplitude is
\begin{eqnarray}
{\cal A}_{\rm SM}(q_L\overline{q_L}\rightarrow \ell_R\overline{
\ell_R})=ig^2\frac{t}{s}{\cal F}_{RL} \;,
\end{eqnarray}
where
\begin{eqnarray}
{\cal F}_{\alpha\beta}=2Q_{\ell}Q_q\sin^2 \theta_{\rm
W}+\frac{s}{s-m^2_{\rm
Z}}\frac{2g^{\ell}_{\alpha}g^{q}_{\beta}}{\cos^2 \theta_{\rm w}} \;.
\end{eqnarray}
 Here $g_L^f= T_{3f} - Q_f \sin^2\theta_{\rm w}$, $g_R^f = - Q_f 
\sin^2\theta_{\rm
 w}$, $Q_f$ is the electric charge of the fermion $f$ in units of proton
charge, and the $SU(2)_L$ coupling $g=e/{\sin \theta_{\rm w}}$.
Identifying the string coupling with the gauge coupling $g_S=g$
and matching the Chan-Paton factor $T_{ijkl}$ as
\begin{eqnarray}
T_{1243}=T_{1324}\equiv T, && T_{1234}=T+1
\end{eqnarray}
one can demand the string amplitude expression in Eq.(\ref{str}) to reproduce
the standard model amplitude in the low-energy limit. Since the
string amplitude describes massless particle scattering, the
masses of the $W$ and $Z$ gauge bosons are introduced by hand. So
Eq. (\ref{str}) becomes
\begin{eqnarray}
\label{strlow}
 &{\cal A}_{\it string}(q_L\overline{q_L}\rightarrow \ell_R\overline{\ell_R})
 =ig^2S(s,t)\frac{t}{s}{\cal F}_{RL}+ig^2T\frac{t}{us}f(s,t,u) \;,\\
&f(s,t,u)\equiv uS(s,t)+sS(t,u)+ tS(u,s) \;.
\end{eqnarray}
The amplitude for $q_R\overline{ q_R}\rightarrow
\ell_L\overline{\ell_L}$ process is obtained from Eq.
(\ref{strlow}) by changing ${\cal F}_{RL}$ to ${\cal F}_{LR}$. As
for the $q_R\overline{ q_R}\rightarrow \ell_R\overline{\ell_R}$
and $q_L\overline{ q_L}\rightarrow \ell_L\overline{\ell_L}$
processes, we use the corresponding ${\cal F}_{RR}$ and ${\cal
F}_{LL}$, respectively, and interchange $t\leftrightarrow u$ in
Eq. (\ref{strlow}).

The factor $\Gamma(1-s/{M_S}^2)$ in the Veneziano amplitude
$S(s,t)$ develops poles at $s=n M_S^2 (n=1,2,3...)$, implying
resonance states with masses $\sqrt{n}M_S$. In the limit $M_S \gg
\sqrt{s}, \sqrt{|t|}, \sqrt{|u|}$
\begin{eqnarray}
&&S(s,t)\approx 1-\frac{\pi^2}{6}\frac{st}{M_S^4},
 S(t,u)\approx1-\frac{\pi^2}{6}\frac{tu}{M_S^4},
 S(u,s)\approx 1-\frac{\pi^2}{6}\frac{us}{M_S^4} \label{Sapp} \;, \\
&& f(s,t,u)\approx -\frac{\pi^2}{2}\frac{stu}{M_S^4}\label{f} \;.
\end{eqnarray}
Therefore the amplitude squared for Drell-Yan process is given by
\[
\sum |{\cal M}|^2 = 4 u^2 \left( |M^{\ell q}_{LL}(s)|^2 + |M^{\ell
q}_{RR}(s)|^2 \right ) + 4 t^2 \left( |M^{\ell q}_{LR}(s)|^2 +
|M^{\ell q}_{RL}(s)|^2 \right ) \;,
\]
where
\begin{equation}
\label{mab1} M^{\ell q}_{\alpha\alpha}(s) = e^2 \Biggl [
\left(\frac{Q_\ell Q_q}{s} + \frac{g_\alpha^\ell
g_\alpha^q}{\sin^2\theta_W \cos^2 \theta_W } \frac{1}{s - M_Z^2
}\right)\left( 1- \frac{\pi^2su}{6M_S^4}\right)-
\frac{\pi^2uT}{4\sin^2\theta_WM_S^4}\Biggr ],\; \alpha=L,R
\end{equation}
\begin{equation}
\label{mab2} M^{\ell q}_{\alpha\beta}(s) = e^2 \Biggl [
\left(\frac{Q_\ell Q_q}{s}+ \frac{g_\alpha^\ell
g_\beta^q}{\sin^2\theta_W \cos^2 \theta_W } \frac{1}{s - M_Z^2
}\right)\left( 1- \frac{\pi^2st}{6M_S^4}\right)-
\frac{\pi^2tT}{4\sin^2\theta_WM_S^4}\Biggr ]
\begin{array}{c}
\alpha,\beta=L,R\\
\alpha\neq\beta
\end{array}
\end{equation}
 Based on the above formulas the square of amplitudes for other
processes can be obtained by crossing the Mandelstam variables and
replacing the corresponding gauge boson.

\section{High Energy Processes and Data Sets}
\label{sec:data}

Before describing the data sets used in our analysis, let us first
specify certain important aspects of the analysis technique. Since
the next-to-leading order calculations do not exist for the new
interaction, we use leading order (LO) calculations for
contributions from both the SM and the new interactions, for
consistency. However, in many cases, e.g. in the analysis of
precision electroweak parameters, it is important to use the best
available calculations of their SM values, as in many cases data
is sensitive to the next-to-leading and sometimes even to
higher-order corrections. Therefore, we normalize our leading
order calculations to either the best calculations available, or
to the low-$Q^2$ region of the data set, where the contribution
from the string resonances is expected to be vanishing. This is
equivalent to introducing a $Q^2$-dependent $K$-factor and using
the same $K$-factor for both the SM  contribution and the effects
of string resonances. The details of this procedure for each
data set are given in the corresponding section. Wherever parton
distribution functions (PDFs) are needed, we use the CTEQ5L
(leading order fit) set \cite{cteq5}. The reason to use the LO PDF
set is that LO PDFs are extracted using LO cross section
calculations, thus making them consistent with our approach.

\subsection{Drell-Yan Production at the Tevatron}

Both CDF \cite{dy-cdf} and D\O\ \cite{dy-d0} measured the
differential cross section $d\sigma/dM_{\ell\ell}$ for Drell-Yan
production, where $M_{\ell\ell}$ is the invariant mass of the
lepton pair. (CDF analyzed data in both the electron and muon
channels; D\O\ analyzed only the electron channel.) The
differential cross section, including the contributions from the
string resonances, is given by
$$
        \frac{d^2\sigma}{dM_{\ell\ell} dy} = K
\frac{M_{\ell\ell}^3}{72\pi s}
        \;
        \sum_q f_q(x_1) f_{\bar q}(x_2)\; \left( |M^{\ell q}_{LL}(\hat
s)|^2 +
        |M^{\ell q}_{LR}(\hat s)|^2 + |M^{\ell q}_{RL}(\hat s)|^2 +
|M^{\ell q}_{RR}(\hat
        s)|^2 \right ) \;,
$$
where $M_{\alpha\beta}^{\ell q}$ is given by Eqs. (\ref{mab1}) and
(\ref{mab2}), $\hat s=M^2_{\ell\ell}$, $\sqrt{s}$ is the
center-of-mass energy in the $p\bar p$ collisions, $M_{\ell\ell}$
and $y$ are the invariant mass and the rapidity of the lepton
pair, respectively, and $x_{1,2} = \frac{M_{\ell\ell}}{\sqrt s}
e^{\pm y}$. The variable $y$ is integrated numerically to obtain
the invariant mass spectrum.   The QCD $K$-factor is given by $K=
1+\frac{\alpha_s(\hat s)} {2 \pi} \frac{4}{3} (1 +
\frac{4\pi^2}{3})$. We scale this tree-level SM cross section by
normalizing it to the $Z$-peak cross section measured with the
data by a scale factor $C$ ($C$ is very close to 1 numerically).
The cross section $\sigma$ used in the fitting procedure is given
by
\begin{equation}
        \label{sigma}
        \sigma = C \left( \sigma_{\rm SM} + \sigma_{\rm interf} +
        \sigma_{\rm string} \right )\; ,
\end{equation} where $\sigma_{\rm interf}$ is the interference term
between the SM and the string resonance amplitudes and
$\sigma_{\rm string}$ is the cross section due to the string
resonances interactions only. When normalizing to the low-energy
data, we neglect the possible contribution from the string
resonances, as it is much smaller than the experimental
uncertainty on the data that we use.

\subsection{HERA Neutral and Charged Current Data}

ZEUS~\cite{zeusccp,zeusncp} and H1 \cite{H1ncccp} have published
results on the neutral-current (NC) and charged-current (CC)
deep-inelastic scattering (DIS) in $e^+ p$ collisions at $\sqrt{s}
\approx 319$ and $318$ GeV, respectively. The data sets collected
by H1 and ZEUS correspond to an integrated luminosities of 65.2
(H1), 60.9 (ZEUS CC) and 63.2 (ZEUS NC) pb$^{-1}$.
ZEUS~\cite{zeusccn,zeusncn} and H1~\cite{H1ncccn} have also
published NC and CC analysis for the data collected in $e^- p$
collisions at $\sqrt{s} \approx 318$ and $320 $ GeV, respectively,
with an integrated luminosity of 16.4 (ZEUS CC), 15.9 (ZEUS NC),
and 16.4 (H1) pb$^{-1}$.

We used the single differential cross section $d\sigma/d Q^2$
presented by ZEUS \cite{zeusccp,zeusncp} and double differential
cross section $d^2 \sigma/dx dQ^2$ published by H1 \cite{H1ncccp}.
The double differential cross section for NC DIS in the $e^+ p$
collisions, including the effect of the string resonances, is
given by
\begin{eqnarray}
        &&\frac{d^2\sigma}{dx dQ^2} (e^+p\to e^+X) = \nonumber \\
        &&  \frac{1}{16\pi}\; \Biggr \{
        \sum_q f_q(x) \,\biggr [ (1-y)^2 ( |M^{eq}_{LL}(t)|^2 +
        |M^{eq}_{RR}(t)|^2) + |M^{eq}_{LR}(t)|^2 + |M^{eq}_{RL}(t)|^2
\biggr ] \label{nccc}
        \\
        && + \sum_{\bar q} f_{\bar q} (x) \, \biggr [
|M^{eq}_{LL}(t)|^2 +
        |M^{eq}_{RR}(t)|^2 + (1-y)^2 (|M^{eq}_{LR}(t)|^2 +
|M^{eq}_{RL}(t)|^2)
        \biggr ] \; \Biggr \} \;, \nonumber \end{eqnarray} where $Q^2
= sxy$ is the square of the momentum transfer and $f_{q/\bar
q}(x)$ are parton distribution functions. The reduced amplitudes
$M_{\alpha\beta}^{eq}$ are given by Eqs. (\ref{mab1}) and
(\ref{mab2}). The double differential cross section for CC DIS,
including the effect of string resonances, can be written as
\begin{eqnarray}
        \label{cccc}
        \frac{d^2\sigma}{dx dQ^2} (e^+p\to \bar \nu X)\, =\,
        \frac{g^4}{64\pi}\biggr\{
        \; &\left | \frac{1}{-Q^2-M_W^2}\left( 1-\frac{\pi^2tu}{6M_S^4}\right)
         - \frac{\pi^2uT}{2M_S^4} \right |^2&
        \; \biggr [ (1-y)^2 ( d(x) + s(x))\biggr]\nonumber\\+
        &\left | \frac{1}{-Q^2-M_W^2}\left( 1-\frac{\pi^2st}{6M_S^4}\right)
         - \frac{\pi^2sT}{2M_S^4} \right |^2&
        \; \biggr [\bar u(x) + \bar c(x) \biggr ]\biggr\} \;,
        \;
\end{eqnarray}
 where $d(x),s(x),\bar u(x), \bar c(x)$ are the parton distribution
functions.  The single differential cross section $d\sigma/d Q^2$
is obtained from the above equations by integrating over $x$. The
cross section in the $e^- p$ collisions can be obtained by
interchanging $( LL \leftrightarrow LR, RR \leftrightarrow RL)$ in
Eq. (\ref{nccc}) and by interchanging $(q(x) \leftrightarrow \bar
q(x) )$ in Eq. (\ref{cccc}).

We normalize the tree-level SM cross section to that measured in
the low-$Q^2$ ($Q^2 \leq 2000$ GeV$^2$) region. The cross section used in
the fitting procedure is then obtained similarly to that in Eq.
(\ref{sigma}).

\subsection{LEP~2 Data}
We analyze the LEP~2 observables sensitive to the effects of the
string resonances, including hadronic and leptonic cross sections.
The LEP Electroweak Working Group combined the $q \bar q, \mu^+
\mu^-$, and $\tau^+ \tau^-$ data from all four LEP collaborations
\cite{lepew} for the machine energies between 130 and 207 GeV. We
use the following quantities in our analysis: (i) total hadronic
cross sections, (ii) total $\mu^+ \mu^-$, $\tau^+ \tau^-$ cross
sections. We take into account the correlations of the data points
in each data set as given by~\cite{lepew}. For Bhabha scattering
cross section $\sigma(e^+ e^- \to e^+ e^-)$, we use various data
sets from individual experiments~\cite{delphi,opal}.

The angular distribution for $e^- e^+ \to f \bar f\;
(f=q,e,\mu,\tau)$ is given by
\begin{eqnarray}
        &&\frac{d\sigma}{d\cos\theta} = \nonumber \\
        &&\frac{N_f s}{128 \pi} \, \Bigg\{
        (1+\cos\theta)^2 \, \left( |M^{ef}_{LL}(s)|^2 +
|M^{ef}_{RR}(s) |^2
        \right ) +(1-\cos\theta)^2 \, \left( |M^{ef}_{LR}(s)|^2 +
        |M^{ef}_{RL}(s) |^2 \right ) \nonumber \\
        && +\delta_{ef} \biggr[(1+\cos\theta)^2 \,\left(
|M^{ef}_{LL}(s)+
        M^{ef}_{LL}(t)|^2 + |M^{ef}_{RR}(s)+M^{ef}_{RR}(t)|^2 -
        |M^{ef}_{LL}(s)|^2 - |M^{ef}_{RR}(s) |^2 \right ) \nonumber \\
        && +4 \left( |M^{ef}_{LR}(t)|^2 + |M^{ef}_{RL}(t) |^2 \right
        ) \biggr] \; \Biggr \} \;,\nonumber
\end{eqnarray} where $N_f=1$ (3) for $\ell$ $(q)$, and $M_{\alpha\beta}^{ef}$
 is given by Eqs. (\ref{mab1}) and (\ref{mab2}).  The additional
terms for $f=e$ arise from the $t$,$u$-channel exchange diagrams.

To minimize the uncertainties from higher-order corrections, we
normalize the tree-level SM calculations to the NLO cross section,
quoted in the corresponding experimental papers. We then scale our
tree-level results, including contributions from the $Z$,
$\gamma$, and string resonances with this normalization factor,
similar to Eq. (\ref{sigma}). When fitting angular distribution,
we fit to the shape only, and treat the normalization as a free
parameter of the fit.

\section{Constraints from High Energy Experiments}
\label{sec:results}

In the previous section, we have described the data sets from
various high energy experiments used in our analysis.
Based on the above individual and combined data sets, we perform a
fit to the sum of the SM prediction and the contribution of the
string resonances, normalizing our tree-level cross section to the
best available higher-order calculations, as explained above. As
seen from Eqs. (\ref{mab1}) and (\ref{mab2}), the effects of the
string resonances always enter the equations in the form
$1/M_S^4$. Therefore, we parameterize these effects with a single
fit parameter $\eta$:
$$
        \eta \equiv\frac{1}{M_S^4} \;.
$$
In most cases, the differential cross sections in presence of the
string resonances are bilinear in $\eta$. We use MINUIT to
minimize the $\chi^2$.

\begin{figure}[t!]
\begin{center}
\includegraphics[width=5in]{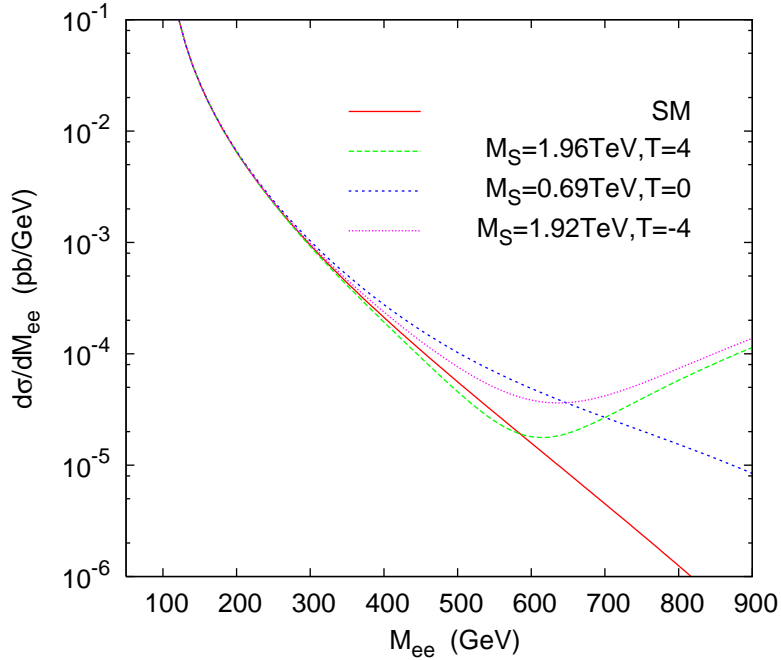}
\caption{\small The differential distribution $d\sigma/dM_{ee}$ for
Drell-Yan production at the Tevatron (Run I) for the low scale open-string
model and for the SM. \label{fig1}} 
\end{center}
\end{figure}

The best-fit values of $\eta$ for each individual data set and
their combinations are shown in Table \ref{table1}. In all cases,
the preferred values from the fit are consistent with zero, and
therefore we proceed with setting limits on $\eta$. The one-sided
95\% CL upper limit on $\eta$ is defined as: \begin{equation}
        0.95 = \frac{\int_0^{\eta_{95}} d \eta \; P(\eta) }
        {\int_0^\infty d \eta \; P(\eta) } \;,\label{maxL}
\end{equation}
where $P(\eta)$ is the fit likelihood function given by
\begin{equation}
 P(\eta)=\frac{1}{\sigma\sqrt{2\pi}}
\exp( -(\chi^2(\eta) - \chi^2_{\rm min})/2 ) \;.
\end{equation}
The corresponding upper 95\% CL limits on $\eta$ and lower 95\%
CL limits on $M_S$ are also shown in Table \ref{table1}. The
combined limit is as high as 1.96 TeV (1.92) TeV for $T=4 (-4)$.
Note that even for $T=0$, there is still some stringy effects
$(1/M_S^4)$, as shown in Eqs. (\ref{mab1}) and (\ref{mab2}). It is
clear from the results of Table~\ref{table1} that the dominant set
of data affecting the limit on $M_S$ is the Drell-Yan data. This
is simply because the effect of string resonances scales like
$1/M_S^4$. We show the effect of string resonances on the
invariant mass spectrum in Fig.~\ref{fig1}.

\section{Sensitivity at the LHC}
\label{sec:sensitivity}
 We expect that at the LHC the best channel
to probe the string resonances is Drell-Yan production. We assume
that the approximation $M_S^2 \gg \hat s, |\hat t|, |\hat u|$ is
still valid at the LHC. Therefore, the reduced amplitudes of Eqs.
(\ref{mab1}) and (\ref{mab2}) can be used and tested by a direct
comparison at the LHC.

It was shown in Ref.\cite{greg} that using the double differential
distribution $d^2\sigma/ M_{\ell\ell} d \cos\theta$ can increase
the sensitivity to the KK states of the graviton compared to the
use of single differential distributions. Similarly, we expect
this to be the case for the string resonances. The double
differential cross section for Drell-Yan production, including the
interactions of the $\gamma$, $Z$, and string resonances, is
given by
\begin{eqnarray}
        \frac{d^3\sigma}{dM_{\ell\ell} dy d\cos\theta^*} &=& K \sum_q
        \frac{M_{\ell\ell}^3}{192 \pi s} f_q(x_1) f_{\bar q}(x_2)
        \, \biggr[ (1+\cos\theta^*)^2 \left( |M_{LL}^{eq}(\hat s)|^2
        +|M_{RR}^{eq}(\hat s)|^2 \right) \nonumber \\
        &&+ (1-\cos\theta^*)^2
        \left( |M_{LR}^{eq}(\hat s)|^2 +|M_{RL}^{eq}(\hat s)|^2
\right)
        \biggr ] \;,\nonumber
\end{eqnarray} where $M_{\alpha\beta}^{eq}$ are given by Eqs. (\ref{mab1})
 and (\ref{mab2}), $\theta^*$ is the scattering
angle in the center-of-mass frame of the initial partons, $\hat s
= M_{\ell\ell}^2$, $dx_1 dx_2 = (2 M_{\ell\ell}/s) dM_{\ell\ell}
dy$, and $x_{1,2} = M_{\ell\ell} e^{\pm y}/\sqrt{s}$.

We briefly describe the procedures here. (i) We calculate the SM
cross section in each bin. (ii) In each bin, we obtain the
expected number of events by multiplying the SM cross section
by the integrated luminosity. (iii) In each bin, we
generate the number of events according to the expect number of
events. If the expected number of events is less than 100, we used
the Poisson statistics otherwise we used the Gaussian. (iv) After
generating the data set, we fit the data set to $\eta$ using Eq.
(\ref{sigma}). Then we evaluate the 95\% CL lower limit on $M_S$
implied by the data. (v) We repeat the above procedure for 1000
times. Then we histogram the limit obtaining each time. We take
the sensitivity to be the medium (501st) of the histogram.

In each simulation, we always normalize to the $Z$-peak data, by
which a lot of systematic uncertainties are eliminated. We have
used an efficiency of 90\% and a rapidity coverage of $|\eta| <
2.0$ for lepton detection. 
We also used a single experiment with combined $e, \mu$
samples. The sensitivity may be lowered by about 10\% if some
level of systematic uncertainties are used in calculating the
$\chi^2$ \cite{greg}. 
The sensitivity, at the 95\% CL, to $M_S$ at the LHC
(100 fb$^{-1}$) with different Chan-Paton factors is given in
Table~\ref{table2}. The sensitivity reach is as high as 19 TeV for
$|T|=4$.
Note that the sensitivity obtained in Ref. \cite{t.han} is somewhat 
lower than the sensitivity obtained in this paper, because their 
analysis is based on the search for a single resonance state while 
ours is based on the deviation in the whole invariant mass spectrum.
Therefore, statistically we can obtain a higher sensitivity.  

\section{Conclusions}
\label{sec:summary} 
The ultimate consequence of a low string scale
is the testable string scattering amplitudes at hadron collider.
The scattering of particles shows the stringy nature. The most
obvious sign is the presence of string resonances. Even if the
energy is not high enough, some deviations from the SM due to the
string resonances are still expected. For example, the tree-level
open-string amplitude for dilepton production at low
energy can be expressed generically as
 \[
 {\cal A}_{\it string} \sim {\cal A}_{\it
 SM}(s,t,u)\cdot S(s,t,u)+Tf(s,t,u)\cdot g(s,t,u) \;,
 \]
 where ${\cal A}_{\it SM}$ is the SM amplitudes, $S(s,t,u)$ are the
 Veneziano amplitudes given in Eq. (\ref{Sapp}), $T$ is the undetermined
 Chan-Paton parameter,$f(s,t,u)$ a kinematic function given in 
Eq. (\ref{f}), and $g(s,t,u)$
 some process-dependent kinematic functions. In the low energy
 limit $s \ll M_S^2\;, S(s,t) \rightarrow 1,\; f(s,t,u) \rightarrow 0$,
the open-string amplitude reduces to the standard model amplitude.

In this work we have performed a global analysis of the model
based on the data from Drell-Yan production at the Tevatron, HERA
neutral and charged current deep-inelastic scattering, and fermion
pair production at LEP~2. Drell-Yan production at the Tevatron,
due to large invariant mass data, provides the strongest
constraint. Combining all data the string scale $M_S$ must be
$\geq 0.7 - 2$ TeV for Chan-Paton factors $|T|=0 - 4$ at 95\% CL.
We have also estimated the expected sensitivity at the LHC, which
is about 19 TeV for $|T|=4$.

\section*{Acknowledgments}
This
research was supported in part by the National Science Council of Taiwan
R.~O.~C.\ under Grant Nos.\ NSC 93-2112-M-007-025-.

\begin{table}[t!]
\caption{\small Best-fit values of $\eta=1/M_S^4$ and the 95\% CL upper
limits on $\eta$ for individual data set and combinations.
Corresponding 95\% CL lower limits on $M_S$ are also shown. By
default we use $T=4$ in part(a). We show in (b) limits for all
combined data with different $T$. \label{table1} }
\medskip 
\begin{ruledtabular}
\begin{tabular}{lccc}
\multicolumn{4}{c}{(a)} \\
   & $\eta$ (TeV$^{-4}$)  & $\eta_{95}$ (TeV$^{-4}$) &  $M_S^{95}$
(TeV) \\
\hline \hline
TEVATRON:              & & & \\
{} Drell-yan           & $-0.013^{+0.070}_{-0.044}$ & 0.074 & 1.92 \\
\hline
HERA:     & & & \\
{} NC     & $0.311^{+0.208}_{-0.192}$ & 0.706 & 1.09 \\
{} CC     & $1.318^{+1.186}_{-1.223}$ & 3.33 & 0.74 \\
{} HERA combined & $0.336^{+0.210}_{-0.192}$ & 0.73  & 1.08 \\
\hline
LEP~2:                       & & & \\
{} hadronic cross section \& ang. dist.
    & $-1.17^{+0.34}_{-0.28}$ & 2.92  & 0.76 \\
{} $ee,\mu\mu,\tau\tau$ cross section \& ang. dist.
      & $-0.122^{+0.097}_{-0.098}$ & 0.12 & 1.68 \\
{} LEP~2 combined          & $-0.156^{+0.098}_{-0.098}$ &0.11&1.72  \\
\hline
All combined (T=4) & $-0.023^{+0.059}_{-0.034}$ & 0.066 &1.96 \\
\end{tabular} 

\begin{tabular}{lccc}
\multicolumn{4}{c}{(b)} \\
 \hline
All combined (T=1) & $-0.044^{+0.227}_{-0.166}$ & 0.27 & 1.38 \\
All combined (T=0) & $2.28^{+1.27}_{-1.27}$ & 4.38 & 0.69 \\
All combined (T=$-1$)& $0.138^{+0.114}_{-0.193}$ & 0.29 & 1.35\\
All combined (T=$-4$)& $0.028^{+0.031}_{-0.055}$ & 0.072 & 1.92 \\
\end{tabular}
\end{ruledtabular}
\end{table}

\begin{table}[t!]
\caption{ Sensitivity to the parameter $\eta=1/M_S^4$ at the LHC,
using the dilepton channels. The corresponding 95\% CL lower limits
on $M_S$ are also shown. \label{table2} }
\medskip
\begin{ruledtabular}
\begin{tabular}{ccc}
LHC (14 TeV, 100 fb$^{-1}$) Dilepton & $\eta_{95}$ (TeV$^{-4}$) &
95\% CL lower limit on $M_S$ (TeV)
\\
\hline 
{}T=4 & $8.49\times10^{-6}$ & 18.5 \\
{}T=1 & $3.71\times10^{-5}$ & 12.8 \\
{}T=0 & $5.79\times10^{-4}$ &6.4 \\
{}T=$-1$ & $3.42\times10^{-5}$ & 13.0 \\
{}T=$-4$ & $7.95\times10^{-6}$ & 18.8 \\
\end{tabular}
\end{ruledtabular}
\end{table}

\end{document}